\documentclass[pmlr]{jmlr}

\usepackage{booktabs} 
\usepackage{graphicx} 
\usepackage{url}      
\usepackage{amsmath}  
\usepackage{algorithm2e} 

\newcommand\independent{\protect\mathpalette{\protect\independenT}{\perp}}
\def\independenT#1#2{\mathrel{\rlap{$#1#2$}\mkern2mu{#1#2}}}
\newcommand{\abs}[1]{\left| #1 \right|}
\def\pen{{ \mathrm{pen} }}
\def\Pen{{ \mathrm{Pen} }}
\def\tr{{ \mathrm{tr} }}
\def\diag{{ \mathrm{diag} }}
\DeclareMathOperator*{\argmax}{argmax}

\jmlrvolume{TBD}
\jmlryear{2026}
\jmlrworkshop{Probabilistic Graphical Models (PGM)}

\title{Parameterising Gaussian Graphical Models}

\author{\Name{Jack Storror Carter} \Email{jack.carter@upf.edu}\\
  \addr Department of Business and Economics \\
  Universitat Pompeu Fabra \\ 
  Barcelona, Spain
}

\editor{Gustau Camps-Valls, Manuele Leonelli and Gherardo Varando}

\begin{document}

\maketitle

\begin{abstract}
Gaussian graphical models (GGMs) describe the dependence structure among jointly Gaussian random variables. However, the most common parameterisation of GGMs, the precision matrix, describes both the dependence and scale of the variables. This has been shown to lead to model selection methods that depend on the scale of the variables, despite graphical models being scale invariant. Even after standardising data to have unit sample variances, entries of the precision matrix can be on different scales leading to poor model selection. This paper decomposes the precision matrix into marginal variances and interpretable scale-invariant parameters - the partial correlations and variance inflation factors. This decomposition gives new insights into the precision matrix entries and the dynamics of model selection methods such as penalised likelihoods. In particular, it explains the observed phenomenon that penalties on the precision matrix perform poorly at selecting hub variables and motivates the necessity of data standardisation. It also shows why methods based on partial correlations have better hub detection properties. The effect of penalisation of different quantities on the estimation of marginal variances is then investigated and an interesting simplification of the log-likelihood is shown when using maximum likelihood estimation of the marginal variances.
\end{abstract}

\begin{keywords}
Gaussian graphical model, precision matrix, correlation matrix, partial correlation, model selection, penalised likelihood.
\end{keywords}

\section{Introduction}

The distribution of a Gaussian random vector $X=(X_1,\dots,X_p)$ is fully specified by its first and second order moments. The univariate moments correspond to the marginal properties of location, through the first order moments $\mathbb{E}[X_i]$, and scale, through the second order moments $\mathbb{E}[X_i^2]$. The bivariate second order moments $\mathbb{E}[X_iX_j]$ describe the dependence structure between variables - in the case of Gaussian random vectors, dependence is completely determined by these pairwise relationships. However, the standard Gaussian parameterisation of the mean vector $\mu=0$, which is assumed equal to 0 for simplicity, and the positive definite covariance matrix $\Sigma$ does not cleanly separate the concepts of scale and dependence. This is because the off-diagonals of $\Sigma$ are a function of both the marginal variances and the pairwise correlations. For this reason a popular alternative is to parameterise directly by the marginal variances and the correlation matrix. The correlation matrix is scale invariant and therefore gives a clean description of the dependence structure.

Gaussian graphical models (GGMs) combine the Gaussian distributional assumption with the conditional independence assumptions of an undirected graphical model (see \cite{uhler2018gaussian} for an overview). A graphical model is a model of only the dependence structure of $X$ and is invariant to scale. As such, a graphical model only makes assumptions about the correlation matrix. However, the connection between marginal correlations and conditional independence is non-trivial which lead to alternative parameterisations.

The most popular parameterisation for GGMs is the precision matrix $\Theta := \Sigma^{-1}$ due to its connection to conditional independence - $\Theta_{ij}=0$ if and only if $X_i \independent X_j \mid X_{-(i,j)}$. The sparsity structure of $\Theta$ therefore directly implies a graphical model. However, while the zeros in $\Theta$ indicate conditional independence, the non-zeros are not a good measure for conditional dependence leading to two main issues. First, being a one to one transformation of $\Sigma$ rather than the correlation matrix, $\Theta$ is not scale invariant. The issue is more pronounced here because, unlike $\Sigma$ where only the off-diagonals conflate scale and dependence, every entry of $\Theta$ is a function of both scale and dependence. Second, as will be demonstrated in Section \ref{sec:PrePC}, even when data has been standardised, the non-zero off-diagonals of $\Theta$ are not easy to interpret and are not a clear measure of strength of conditional dependence.

These issues were partly addressed by \cite{carter2024partial} who pointed out the lack of scale invariance of $\Theta$ and instead advocated for encoding conditional independence by partial correlations. Partial correlations are inherently scale invariant and are a clear measure of conditional dependence. However, this only addressed part of the issue with the remaining diagonal entries of $\Theta$ also being functions of both scale and dependence. Therefore, a parameterisation combining partial correlations with the diagonal of $\Theta$ does not fully separate the two concepts. This paper will further decompose the diagonal of $\Theta$ into marginal variances and a scale invariant component - the variance inflation factors (VIFs).

It will be shown that the VIFs are functions of the partial correlations, therefore allowing $\Theta$ to be reparameterised by the pair of marginal variances and partial correlations. This fully separates scale and dependence while also encoding conditional independence. The VIFs also have a clear interpretation as a measure of overall dependence of $X_i$ on $X_{-i}$. Decomposing the entries of $\Theta$ into partial correlations, marginal variances and VIFs gives an interpretable explanation for the precision matrix entries and new understanding on how methods based on the precision matrix work.

The main focus of this paper will be on model selection through the lens of penalised likelihoods. For the precision matrix $\Theta$, this is an estimator of the form
$$\hat \Theta_\Pen(S) := \argmax_{\Theta \succ 0} l(\Theta \mid S) - \sum_{i=1}^p \pen_D(\theta_{ii}) - \sum_{i < j} \pen(\theta_{ij}),$$
where $S$ is the sample covariance matrix, $l$ is the log-likelihood function, $\pen_D$ is a penalty function on the diagonals and $\pen$ is a penalty function on the off-diagonals. Penalised likelihood estimates can analogously be defined for other parameterisations. The penalty function $\pen$ is generally chosen to be decreasing on $(-\infty,0)$ and increasing on $(0,\infty)$ so that estimates are shrunk towards $0$. When the penalty is non-differentiable at $0$, this can result in sparse estimation of $\Theta$.

Many popular methods for GGM selection fit into this framework, for example $l_1$ penalisation via graphical LASSO (GLASSO) \citep{Yuan2007,Banerjee2008,Friedman2008}, non-convex penalties \citep{fan2001variable,fan2009network,zhang2010nearly} and penalties that approximate the $l_0$ \citep{dicker2013variable,wang2016variable,williams2020back}. They are popular due to their simple construction and generally fast computation.

However, any such penalised likelihood method on $\Theta$, other than $l_0$ penalisation, is not scale invariant \citep[Proposition 1]{carter2024partial}, with the resulting estimate and selected graphical model being sensitive to the scaling of the data or standardisation choices. As an alternative, \cite{carter2024partial} proposed direct penalisation of the partial correlations, for which estimation is scale invariant. An $l_1$ penalisation, called partial correlation graphical LASSO (PCGLASSO) was implemented and shown in simulated experiments to outperform GLASSO, even when data was standardised. The advantage of PCGLASSO was most pronounced in hub settings where the data generating graphical model has high degree vertices. This experimental observation was then backed up theoretically by \cite{bogdan2025identifying} who proved model selection consistency of PCGLASSO under a weaker condition than GLASSO, with the greatest gap in these conditions being for hub graphs. Further real data and simulated experiments backed up this theoretical result.

Previously, the poor performance of GLASSO at selecting hubs has been attributed, using a Bayesian argument, to it apriori assigning equal and independent probability to each edge being present, resulting in a small prior probability of high degree nodes \citep{Tan2014}. This motivated an adaptation of the GLASSO penalty to actively encourage hubs \citep{wang2025learning}. However, the improved performance of PCGLASSO in detecting hubs goes against this reasoning. The reparameterisation of $\Theta$ in this paper presents an alternative explanation, showing that GLASSO systematically places higher penalty on hubs, even when data has been standardised. This explanation extends beyond GLASSO to any strictly increasing penalty on $\Theta$, showing that this is an inherent problem of penalised likelihood methods on $\Theta$. On the other hand, penalties on the partial correlations do not have this property of penalising hubs. 

Many other methods have been proposed for GGM selection, with most of them falling into the category of using shrinkage to encourage sparse estimates. For example, estimation of $\Theta$ has been related to a set of linear regressions and some methods use this along with common dimension reducing techniques \citep{meinshausen2006high,liu2015fast,peng2009partial}. Other methods optimise a function other than the log-likelihood, usually combined with an $l_1$ penalty - SPLICE uses a pseudo-likelihood \citep{rocha2008path}, CLIME minimises the $L_1$ norm of $\Theta$ under a constraint that $\Theta^{-1}$ remain close to $S$ \citep{cai2011constrained} and a D-trace loss function has also been proposed \citep{zhang2014sparse}. A fully Bayesian approach to model selection would usually use the conjugate G-Wishart prior \citep{wang2012efficient}. Other Bayesian methods are more similar to the penalised likelihood approach and use a shrinkage prior to encourage sparsity - these include the Bayesian GLASSO \citep{Wang2012}, graphical horseshoe \citep{li2019graphical} and spike and slab priors \citep{Banerjee2015,Gan2018,jewson2024graphical,sulem2025bayesian}. Although the focus of this paper is on penalised likelihood methods, the reparameterisation of $\Theta$ in this paper can be used to give new insights and interpretations in any of these methods.

A number of different parameterisations will be introduced through the paper. For reference, these are summarised in Table \ref{tab:parameters}.

\begin{table}[htbp]
\centering
\label{tab:parameters}
\begin{tabular}{@{}ll@{}}
\toprule
\textbf{Parameter} & \textbf{Description}\\
\midrule
$\Sigma$ & Covariance matrix \\
$\sigma^2 = \mathrm{diag}(\Sigma)$ & Marginal variances \\
$R = \mathrm{cov2cor}(\Sigma)$ & Correlation matrix \\
$\Theta = \Sigma^{-1}$ & Precision matrix \\
$\theta = \mathrm{diag}(\Theta)$ & Precision matrix diagonal (inverse partial variances) \\
$\Delta = \mathrm{cov2cor}(\Theta)$ & Negative partial correlation matrix \\
$\psi = \mathrm{diag}(\Delta^{-1})$ & Variance inflation factors \\
$K = R^{-1}$ & Inverse correlation matrix \\
\bottomrule
\end{tabular}
\caption{Summary of different parameters.}
\end{table}


\section{Covariance and Correlation}\label{sec:CovCor}

The most common parameterisation of a Gaussian distribution is the covariance matrix $\Sigma = (\Sigma_{ij})$. 
The diagonals are marginal variances, $\Sigma_{ii} = \mathrm{Var}(X_i)$, and the off-diagonals are pairwise covariances, $\Sigma_{ij} = \mathrm{Cov}(X_i,X_j)$. For an i.i.d. sample $x_1,\dots,x_n \sim N(0,\Sigma)$ and defining the sample covariance matrix as $S := \frac{1}{n} \sum_{i=1}^n x_i x_i^\top$, the log-likelihood for $\Sigma$, after removing additive and multiplicative constants (as they will be for subsequent log-likelihoods), is
$$ l(\, \Sigma \mid S \,) := -\log\det\Sigma - \tr(S \Sigma^{-1}),$$
and the maximum likelihood estimate (MLE) is $\hat{\Sigma} = S$. The MLE follows a rescaled Wishart distribution $nS \sim \mathcal{W}_p(\Sigma,n)$ and is an unbiased estimate of $\Sigma$. Marginally, the diagonal entry $S_{ii}$ follows a rescaled $\chi^2_n$ distribution with mean $\Sigma_{ii}$ and variance $2\Sigma_{ii}^2/n$ \cite[Section 3.2]{muirhead2009aspects}. Notice that the variance does not depend on the dimension $p$ and so, for moderate sample sizes, one can be reasonably assured that $S_{ii} \approx \Sigma_{ii}$ regardless of dimension.

While the diagonals $\Sigma_{ii}$ describe the scale of the variables and are invariant to dependence between variables, seen by $\mathrm{Var}(X_i)$ being a marginal property, the off-diagonals $\Sigma_{ij}$ are a function of both the scale of and dependence between the variables. That they are not scale invariant follows because scalar multiples of the variables generally have different $\Sigma_{ij}$ - letting $Y_i = a_i + b_iX_i$ with $b_i \neq 0$, $\mathrm{Cov}(Y_i,Y_j) = b_ib_j\mathrm{Cov}(X_i,X_j)$ which are only equal for all $b_i,b_j$ when $\mathrm{Cov}(X_i,X_j)=0$.

To separate the concepts of scale and dependence, one can instead combine the marginal variances, now denoted by the diagonal matrix $\sigma^2=\diag(\sigma_1^2,\dots,\sigma_p^2)$ with $\sigma_i^2 := \Sigma_{ii}$, with the correlation matrix, the matrix $R=(R_{ij})$ with unit diagonal satisfying $\Sigma = \sigma R \sigma$. Using the name of the \texttt{R} function, denote by $R = \mathrm{cov2cor}(\Sigma)$ the function that scales a general positive definite matrix into the corresponding correlation matrix with unit diagonal. Since $\Sigma$ is a one to one function of $(\sigma^2,R)$, this is a valid parameterisation and fully specifies the Gaussian distribution.

The correlations $R_{ij}=\mathrm{Corr}(X_i,X_j)$ are scale invariant because
$$\mathrm{Corr}(Y_i,Y_j) = \frac{\mathrm{Cov}(Y_i,Y_j)}{\sqrt{\mathrm{Var}(Y_i) \mathrm{Var}(Y_j)}} = \frac{b_ib_j\mathrm{Cov}(X_i,X_j)}{\sqrt{b_i^2\mathrm{Var}(X_i) b_j^2\mathrm{Var}(X_j)}} = 
\mathrm{Corr}(X_i,X_j).$$
Hence the $(\sigma^2,R)$ parameterisation cleanly separates the concepts of scale, through $\sigma^2$, and dependence, through $R$. Therefore a graphical model, a model of the conditional independence relationships between variables, only places restrictions on $R$ with no restrictions or assumptions made on $\sigma^2$.


However, because the relationship between correlations and conditional independence is non-trivial, different parameterisations are generally used for GGMs that directly encode conditional independence via sparsity.

\section{Precision and partial correlation}\label{sec:PrePC}

The most popular parameterisation of a GGM is the precision matrix $\Theta := \Sigma^{-1}$. The precision is the canonical parameter of a Gaussian distribution with a particularly simple form for the log-likelihood \citep{uhler2018gaussian}
\begin{equation}\label{eq:PrecisionLikelihood}
    l(\Theta \mid S) := \log \det \Theta - \tr(S\Theta).
\end{equation}
Its diagonal entries are equal to the inverse partial variances $\Theta_{ii} = 1 / \mathrm{Var}(X_i \mid X_{-i})$. The off-diagonals do not have such an obvious interpretation, but the reparameterisations in subsequent sections will provide one.

The popularity of $\Theta$ for parameterising GGMs comes from its clear encoding of conditional independence by its zero entries with $\Theta_{ij} = 0$ if and only if $X_i \independent X_j \mid X_{-(i,j)}$ \citep{uhler2018gaussian}. However, this comes at the cost of harder inference. The MLE of $\Theta$ only exists when $S$ is positive definite, which happens with probability 1 when $n \geq p$ but probability 0 when $n < p$ \cite[Section 8.3]{mathai2022multivariate}.\footnote{This is the case when the mean $\mu$ is known. When $\mu$ is unknown, the sample covariance matrix $S = \frac{1}{n} \sum_{i=1}^n (x_i - \bar x)(x_i - \bar x)^\top$ is positive definite only when $n > p$.} When it exists, the MLE is $\hat \Theta = S^{-1}$ which follows a scaled inverse-Wishart distribution $S^{-1}/n \sim \mathcal{W}^{-1}_p(\Theta,n)$. This is a biased estimate of $\Theta$ with mean $\frac{n}{n-p-1}\Theta$ and the diagonals of $S^{-1}$ follow a scaled inverse-Gamma distribution $(S^{-1})_{ii} / n \sim IG((n-p+1)/2,\Theta_{ii}/2)$ so that $(S^{-1})_{ii}$ has mean $\frac{n}{n-p-1}\Theta_{ii}$ and variance $\frac{2n^2}{(n-p-1)^2(n-p-3)}\Theta_{ii}^2$ \citep[Section 3.3]{gupta1999matrix}. Note that both the bias and variance are increasing with $p$, and the variance is generally much larger than that of $S_{ii}$. Hence one would need a much larger sample size in order to have any reasonable assurance that $(S^{-1})_{ii} \approx \Theta_{ii}$, with the required $n$ increasing with $p$. 

A second issue with the precision matrix is that, being a function of $\Sigma$, it is a measure of both scale and dependence, and the conflation of the two concepts is worse here because every entry of $\Theta$ depends on both scale and dependence. For example, the diagonal of $\Theta$ being the inverse partial variances shows that this is a multivariate property and therefore a function of the dependence between variables. That all entries of $\Theta$ are not scale invariant follows from $\Theta(Y) = b^{-1}\Theta b^{-1}$ where $\Theta(Y)$ is the precision matrix of $Y = a + bX$ and $b = \diag(b_1,\dots,b_p)$.


A first step to decoupling scale and dependence in $\Theta$ is to consider the negative partial correlation matrix, $\Delta := \mathrm{cov2cor}(\Theta)$. The entries of $\Delta$ are $\Delta_{ij} = \Theta_{ij} / \sqrt{\Theta_{ii}\Theta_{jj}} = -\mathrm{Corr}(X_i,X_j \mid X_{-(i,j)})$. That $\Delta$ is scale invariant follows from $\Delta(Y) = \mathrm{cov2cor}(b^{-1}\Theta b^{-1}) = \mathrm{cov2cor}(\Theta) = \Delta$. Writing the diagonal of $\Theta$ as $\theta = \diag(\theta_1,\dots,\theta_p)$ with $\theta_i := \Theta_{ii}$, $\Theta = \theta^{1/2}\Delta\theta^{1/2}$. Hence $(\theta,\Delta)$ is a one to one transformation of $\Theta$ and a valid parameterisation. The log-likelihood is
$$ l(\theta,\Delta \mid S) = \sum_{i=1}^p \log \theta_i + \log \det \Delta - \tr(S \theta^{1/2} \Delta \theta^{1/2}).$$
Since the covariance matrix is related to the correlation matrix via $\Sigma = \sigma R \sigma$ and can also be written in terms of the partial correlation matrix as $\Sigma = (\theta^{1/2}\Delta\theta^{1/2})^{-1} = \theta^{-1/2}\Delta^{-1}\theta^{-1/2}$ it follows that $R = \sigma^{-1} \theta^{-1/2} \Delta^{-1} \theta^{-1/2} \sigma^{-1} = \mathrm{cov2cor}(\Delta^{-1})$, or conversely, $\Delta = \mathrm{cov2cor}(R^{-1})$. This is a one to one transformation meaning that $\Delta$ contains the same information as the correlation matrix and fully determines the dependence structure.

However, under this parameterisation $\theta$ is still a function of both scale and dependence and, as seen above, its estimation is challenging, requiring a large sample size relative to dimension for an accurate MLE.

\section{Variance inflation factors}\label{sec:VRF}

As noted above, the diagonals of the precision matrix are equal to the inverse partial variances $\theta_i = 1/\mathrm{Var}(X_i \mid X_{-i})$. It was seen that this quantity is not scale invariant, and it is also a function of dependence - for fixed marginal variances, $\mathrm{Var}(X_i \mid X_{-i})$ gets smaller as dependence between $X_i$ and $X_{-i}$ increases. To separate scale and dependence, $\theta_i$ can be decomposed into the marginal variance $\sigma_i^2$ and a scale invariant parameter
$$\psi_i := \mathrm{Var}(X_i) / \mathrm{Var}(X_i \mid X_{-i}) = \sigma_i^2 \theta_i,$$ 
which will be referred to as the variance inflation factor (VIF) of $X_i$. This is the factor by which the variance of $X_i$ increases when conditioning on $X_{-i}$ is removed. This is related to the coefficient of determination $R^2$ in linear regression with $\psi_i = 1/(1 - R_i^2)$. It can be shown that $\mathrm{Var}(X_i \mid X_{-i}) \leq \mathrm{Var}(X_i)$ (intuitively, conditioning cannot increase the variance) and therefore $\psi_i \in [1,\infty)$. When $\psi_i=1$, $X_i$ is independent of all of $X_{-i}$, while $\psi_i \to \infty$ means that $X_i$ is a linear function of $X_{-i}$ resulting in a degenerate distribution. A large value of $\psi_i$ implies that the variance of $X_i \mid X_{-i}$ is much smaller than that of $X_i$, meaning that $X_{-i}$ are good predictors of $X_i$. The value of $\psi_i$ therefore gives a measure of the total dependence between $X_i$ and $X_{-i}$. In the graphical model setting, a large $\psi_i$ typically occurs when $X_i$ is of high degree (has many edges) with dependence on many of the $X_{-i}$. However, large $\psi_i$ can also occur when $X_i$ has few edges with particularly strong dependencies.

It can easily be shown that $\psi_i$ is scale invariant $$\psi_i(Y) = \frac{\mathrm{Var}(Y_i)}{\mathrm{Var}(Y_i \mid Y_{-i})} = \frac{b_i^2\mathrm{Var}(X_i)}{b_i^2\mathrm{Var}(X_i \mid X_{-i})} 
= \psi_i.$$
This implies that $\psi_i$ is only a measure of dependence and is therefore a function of $\Delta$, which was shown to fully determine the dependence structure.
Recalling the above connection that $\Delta^{-1} = \theta^{1/2} \sigma R \sigma \theta^{1/2} = \psi^{1/2} R \psi^{1/2}$ where $\psi = \diag(\psi_1,\dots,\psi_p)$, the diagonal of $\Delta^{-1}$ is equal to $\psi$. Similarly, the inverse correlation matrix is $R^{-1} = \sigma \theta^{1/2} \Delta \theta^{1/2} \sigma = \psi^{1/2} \Delta \psi^{1/2}$. So $\psi$ can be calculated directly by the diagonal of either $\Delta^{-1}$ or $R^{-1}$. This also implies that the pair $(\sigma^2,\Delta)$ of marginal variances and partial correlation matrix is a one to one function of $\Sigma$ and so fully determines the Gaussian distribution.

The log-likelihood for $(\sigma^2,\Delta)$ is
$$ l(\sigma^2,\Delta \mid S) = - \sum_{i=1}^p \log \sigma_i^2 + \sum_{i=1}^p \log \psi_i + \log \det \Delta - \tr(S \sigma^{-1} \psi^{1/2} \Delta \psi^{1/2} \sigma^{-1}).$$
This can be simplified by considering the inverse correlation matrix $K := R^{-1} = \psi^{1/2} \Delta \psi^{1/2}$,
\begin{equation}\label{eq:loglikeK}
    l(\sigma^2, K \mid S) = - \sum_{i=1}^p \log \sigma_i^2 + \log \det K - \tr(S \sigma^{-1} K \sigma^{-1}).
\end{equation}
The off-diagonals of $K$ are $\Delta_{ij} \sqrt{\psi_i \psi_j}$ so sparsity still indicates conditional independence.

\section{Effects on model selection}\label{sec:ModelSelec}

The previous sections decomposed the precision matrix into three interpretable parameters: the partial correlations $\Delta_{ij}$ which measure the dependence between $X_i$ and $X_j$ as well as indicating conditional independence; the VIFs $\psi_i$ which measure the total dependence of $X_i$ on $X_{-i}$; and the marginal variances $\sigma_i^2$. The precision matrix off-diagonals can be written in terms of these interpretable parameters as
\begin{equation}\label{eq:decomposition}
    \Theta_{ij} = \frac{\Delta_{ij}\sqrt{\psi_i \psi_j}}{\sigma_i\sigma_j}.
\end{equation}
A penalty function on $\Theta$ therefore has the effect of penalising the partial correlations $\Delta_{ij}$ (the desired effect) as well as penalising the VIFs and inflating the marginal variances (see Section \ref{sec:SigmaEst}). In other words, partial correlations associated to larger VIFs or smaller marginal variances receive larger penalty. In the case of marginal variances, this is bad because the magnitude of penalisation depends on the arbitrary scaling of the variables, however this can be mitigated by data standardisation. In the case of VIFs, this means that variables with the highest overall dependence will receive the most penalisation. When a large $\psi_i$ is due to a small number of strong dependencies, the likelihood is typically strong enough to overpower this high penalty. However, when a large $\psi_i$ is due to a large number of weaker dependencies, i.e. when $X_i$ is a hub, the likelihood function may not have enough evidence for $\Delta_{ij} \neq 0$ and remove these hub edges due to the high penalty associated to them. These phenomena will be demonstrated by an example.

\begin{figure}[h]
\centering
\begin{tabular}{cc}
\includegraphics[scale=0.4]{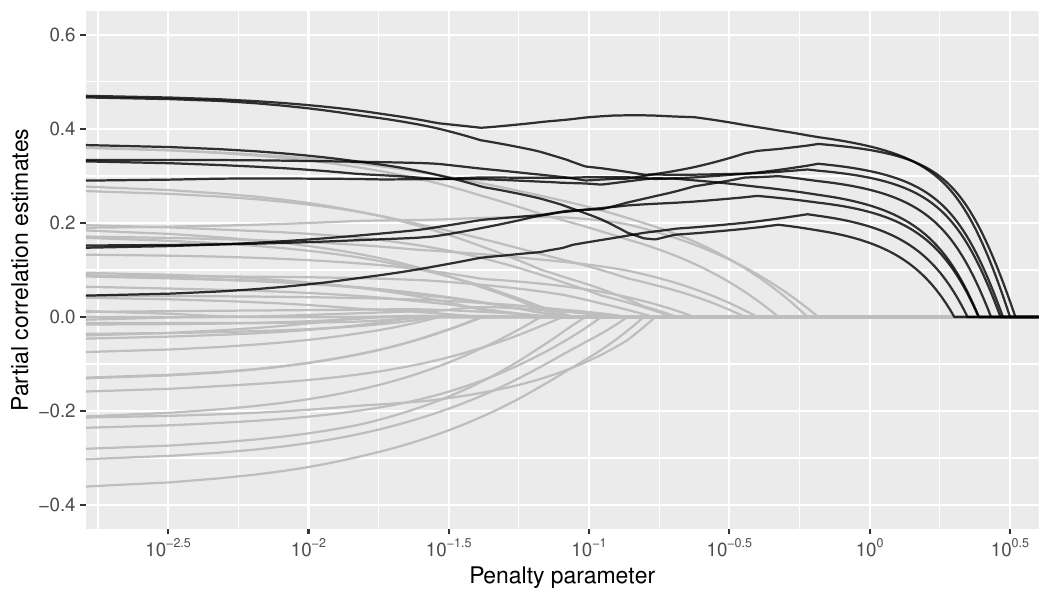} &
\includegraphics[scale=0.4]{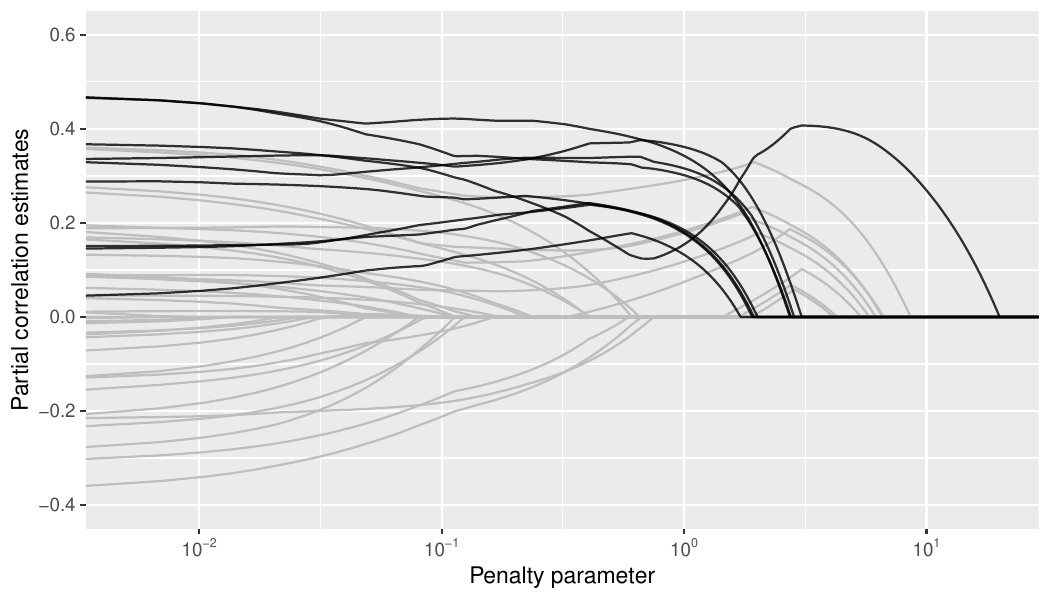} \\
\includegraphics[scale=0.4]{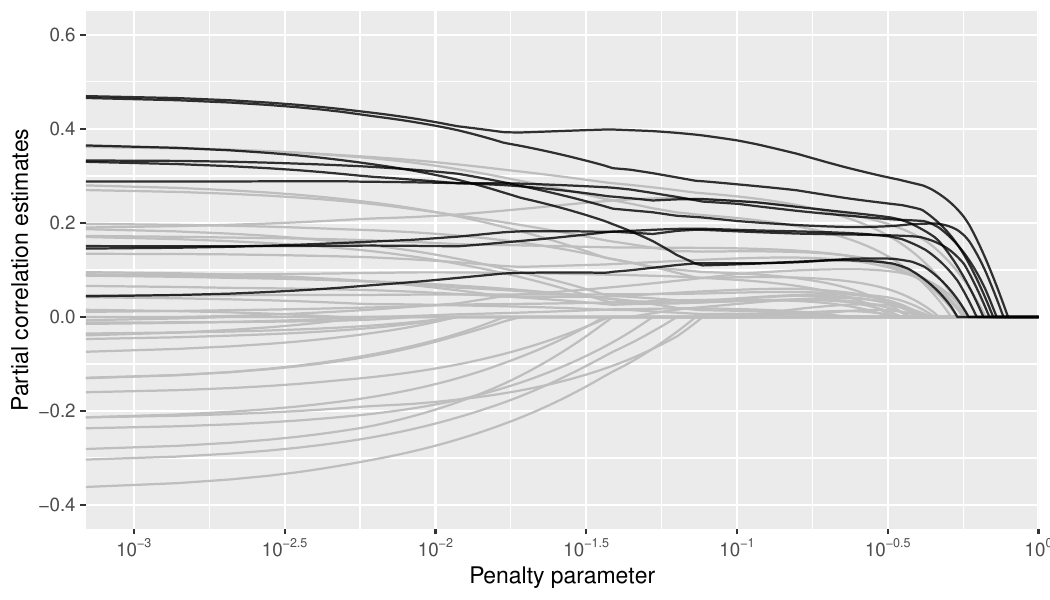} &
\includegraphics[scale=0.4]{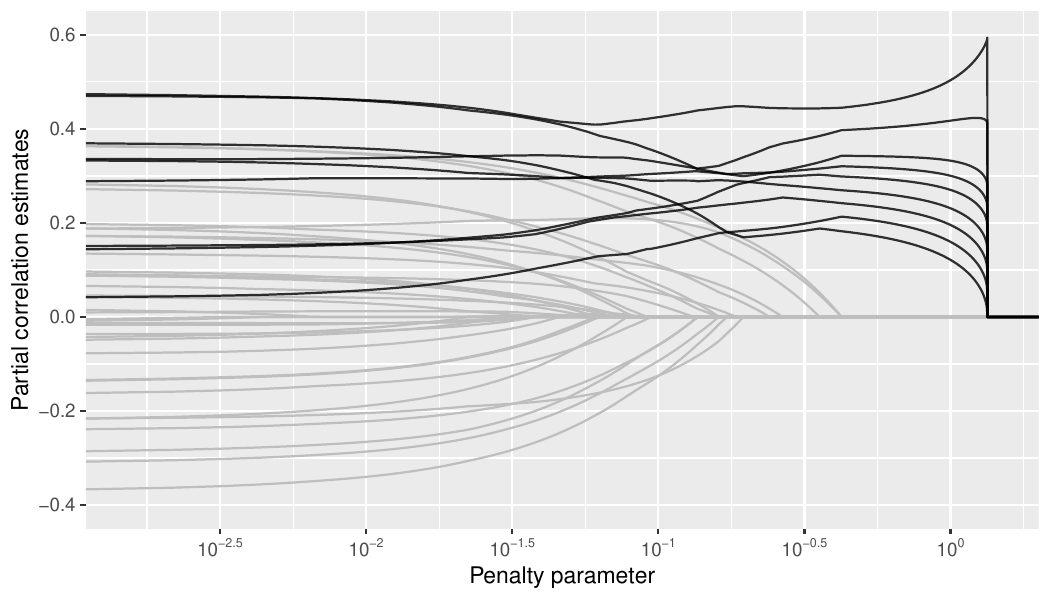} \\
\end{tabular}
\caption{Regularisation paths for hub example. Truly non-zero $\Delta_{ij}$ estimates in black. Top left: GLASSO with $\Theta=\Delta$. Top right: GLASSO with high variance variable. Bottom left: GLASSO on standardised data. Bottom right: PCGLASSO.}
\label{fig:Ex}
\end{figure}

Consider the most extreme hub graph with partial correlations $\Delta_{1j} = -1/\sqrt{p}$, $j=2,\dots,p$ and $\Delta_{ij}=0$ otherwise. For dimension $p=10$, this has $\psi_1=10$ associated to the hub variable and $\psi_i=2$ for all other variables. First consider an idealised setting where $\Theta=\Delta$. Generate $S$ from a sample of size $n=50$ and show the regularisation path for the standard GLASSO (with no diagonal penalty) across different penalty parameters (Figure \ref{fig:Ex}, top left). It is able to recover the true model across quite a large range of penalty parameters. However, if $X_2$ is multiplied by 10, so that now $\theta_2 = 0.01$, then the true model is not recovered for any penalty parameter (Figure \ref{fig:Ex}, top right). The edges associated to $X_1$ are shrunk to 0 before those of $X_2$. Due to $X_2$ having large variance, it receives less penalisation and is selected as a hub variable. To account for variables being on different scales, one should standardise the data such that $S$ has unit diagonal. As discussed in Section \ref{sec:CovCor}, for moderate sample sizes there is some assurance that $\sigma_i^2 \approx 1$ for all $i$ after standardisation. However, now the effect of variables having different $\psi_i$ becomes evident. Again, the true model is not recovered for any penalty parameter (Figure \ref{fig:Ex}, bottom left). In contrast, PCGLASSO is able to select the true model for a large range of penalty parameter values (Figure \ref{fig:Ex}, bottom right) and, being scale invariant, this does not depend on data scales or standardisation.

It is evident from this example that data standardisation is vital when using GLASSO or other similar methods. Without standardisation, high variance variables receive less penalisation and are often selected as hubs, regardless of the evidence for dependence in the likelihood. However, even after standardisation such that $S$ has unit diagonal and $\sigma_i^2 \approx 1$ (in which case $\Theta \approx K$), model selection can suffer when there are large differences in the $\psi_i$ values. These issues have largely gone unnoticed in simulation studies because an idealised setting with $\Theta = \Delta$ is often chosen with no data standardisation. This assumes that $\psi_i = \sigma_i^2$ for all $i$, an unreasonable assumption in any real data.

One might argue that, instead of standardising so that $\sigma_i^2 \approx 1$, one should standardise so that $\theta_i \approx 1$. However, estimation of $\theta_i$ is a fundamentally more difficult task than estimating $\sigma_i^2$ and so such a standardisation is more prone to error. For example, standardising such that $S^{-1}$ has unit diagonal is not possible when $n<p$ and, as pointed out in Section \ref{sec:PrePC}, a large sample size relative to $p$ is required to have any assurance that $(S^{-1})_{ii} \approx \theta_i$. One alternative is an adaptive approach where different penalty parameters are used for each $\Theta_{ij}$ with the penalty scaled based on the currently estimated $\theta$.

A further aspect of penalised likelihood methods is their treatment of the diagonal $\theta$ with shrinkage often also being applied to these parameters. Although not directly indicating model selection through sparsity, the decomposition $\theta_i = \psi_i / \sigma_i^2$ shows that its estimation can still have a great effect on model selection. For fixed $\sigma_i^2$, large $\theta_i$ corresponds to large $\psi_i$ and therefore high dependence between $X_i$ and $X_{-i}$, while small $\theta_i \approx 1 / \sigma_i^2$ indicates near independence between $X_i$ and $X_{-i}$. Shrinkage of $\theta_i$ therefore encourages shrinkage of $\psi_i$ and models with smaller degrees.

\section{Estimation of $\sigma^2$}\label{sec:SigmaEst}

A graphical model is a model of dependence and therefore places no assumptions on the scale parameters $\sigma^2$. One might therefore consider $\sigma^2$ a nuisance parameter in GGM selection. However, estimation of $\sigma^2$ can still play a role in practice since tuning parameters are often learned from the data. If estimation of $\sigma^2$ is poor for certain tuning parameter values, then these are less likely to be chosen regardless of how well they estimate $\Delta$ or the graphical model.

This section proposes two strategies for estimating $\sigma^2$ when treated as a nuisance parameter. It is then shown how penalised likelihood estimates of the various parameterisations actually estimate $\sigma^2$ and whether they correspond to either of these two strategies.

\subsection{Marginal estimation}

An obvious strategy for estimating $\sigma^2$ is to simply fix them equal to the global MLE $\hat \sigma_i^2 = S_{ii}$. This is how the other nuisance parameter, the mean $\mu$, is often treated in practice in GGMs, with its estimate fixed equal to the sample mean and analysis continuing contingent on this estimate. The log-likelihoods for $\Delta$ and $K$ with the fixed $\hat \sigma_i^2 = S_{ii}$ estimate are
\begin{align*}
    l(\Delta \mid S, \sigma^2 &= \diag(S)) = \sum_{i=1}^p \log \psi_i + \log \det \Delta - \tr(C\psi^{1/2}\Delta\psi^{1/2}), \\
    l(K \mid S, \sigma^2 &= \diag(S)) = \log \det K - \tr(CK),
\end{align*}
where $C = \mathrm{cov2cor}(S)$ is the sample correlation matrix. Note that the log-likelihood for $K$ has the same form for the log-likelihood of $\Theta$ in (\ref{eq:PrecisionLikelihood}), but with $S$ replaced by $C$. It therefore seems like estimation of $K$ is analogous to estimation of $\Theta$ with standardised data. However, the difference is that $K$ is an inverse correlation matrix, i.e. $K^{-1}$ has unit diagonal, and so optimisation must be performed on this subspace. This restriction directly implies that the estimated $\sigma^2$ satisfy $\sigma^2 = \diag(S)$, while, as will be seen later, penalised likelihood estimation of $\Theta$ on standardised data can result in vastly different estimates for $\sigma^2$ when the diagonals of $\Theta$ are penalised.

\subsection{Joint estimation}\label{subsec:Joint}

The previous approach of fixing the estimated $\sigma^2$ equal to $\diag(S)$ might seem like the correct approach since it is the same way as the nuisance parameter $\mu$ is dealt with. However, these two things are fundamentally different: the MLE for $\mu$ is always the sample mean, no matter the estimated $\Sigma$, while the MLE for $\sigma^2$ depends on the estimated correlation matrix $R = K^{-1}$. To see this, differentiate the penalised likelihood (\ref{eq:loglikeK}) for $(\sigma^2,K)$ with respect to $\sigma_i$ to arrive at the first order conditions
\begin{equation}\label{eq:sigmaFOCs}
    \sigma_i = \sum_{j=1}^p \frac{S_{ij} K_{ij}}{\sigma_j}, \; i=1,\dots,p,
\end{equation}
This is only satisfied at $\sigma_i^2 = S_{ii}$ when $K = C^{-1}$ (the global MLE), when $K = I_p$ or some linear combination of the two. By writing the optimisation problem in terms of the Hadamard product $S \odot K$, which is guaranteed to be positive definite when $S$ is positive semidefinite with positive diagonal and $K$ is positive definite \citep[Theorem 7.5.3]{horn2012matrix}, it can be shown that the solution to these equations exists and is unique. However, the solution is not generally available in closed form. Denote the solution for a given $K$ or $\Delta$ as $\hat \sigma^2(K)$ and $\hat \sigma^2(\Delta)$.

Since $S_{ij}$ and the global MLE $\hat K_{ij} = (C^{-1})_{ij}$ usually have opposite signs (having the same signs corresponds to a Simpson's paradox setting where conditioning changes the direction of dependence), shrinking $K_{ij}$ away from its MLE $\hat K_{ij}$ towards 0 will generally result in smaller $\hat \sigma^2(K)$. Hence, for penalised likelihood estimates, usually $\hat \sigma^2(K)_i \leq S_{ii}$.

Interestingly, letting $\sigma^2 = \hat \sigma^2(K)$, the trace term in the log-likelihood is constant in $K$
$$ \tr(S \sigma^{-1} K \sigma^{-1}) = \sum_{i=1}^p \sum_{j=1}^p \frac{S_{ij}K_{ij}}{\sigma_i\sigma_j} = p, $$
with the second equality following from the first order condition (\ref{eq:sigmaFOCs}). The log-likelihood in the $K$ and $\Delta$ parameterisations are then
\begin{align*}
    l(K \mid S, \sigma^2 &= \hat \sigma^2(K)) = - \sum_{i=1}^p \log \hat \sigma^2(K)_i + \log \det K, \\
    l(\Delta \mid S, \sigma^2 &= \hat \sigma^2(\Delta) ) = - \sum_{i=1}^p \log \hat \sigma^2(\Delta)_i + \sum_{i=1}^p \log \psi_i + \log \det \Delta.
\end{align*}

\subsection{How shrinkage of different parameterisations affects $\sigma^2$}

Since the decomposition of $\Theta_{ij}$ in (\ref{eq:decomposition}) includes $\sigma_i\sigma_j$ in the denominator, penalisation of $\Delta$ might be expected to inflate the estimates for $\sigma^2$. This is most pronounced when the diagonal of $\Theta$ is penalised. When GLASSO includes the penalty on the diagonal with parameter $\lambda$, the estimated $\sigma_i^2$ is equal to $S_{ii} + \lambda$ \citep{Friedman2008}. The diagonal estimates are similarly inflated when any penalisation is directly applied the diagonal, although the exact form of the inflation depends on the choice of penalty function. These estimates are inflated above the global MLE and are likely even further from the conditional MLE. Hence, for large $\lambda$ the estimated $\sigma_i^2$ do not describe the data well. This means that, when treated as a tuning parameter, large $\lambda$ values will perform relatively poorly resulting in smaller $\lambda$ values, and more dense models, being selected.

When the diagonal penalty is excluded from GLASSO, the estimated marginal variances are exactly equal to $S_{ii}$, the global MLE values. The same can be shown to occur for any penalised likelihood on $\Theta$ which excludes the diagonal penalty \citep{lauritzen2022}.

When the penalty is instead applied only to the partial correlation matrix $\Delta$ (or to the inverse correlation matrix $K$) the penalty is not a function of $\sigma^2$ and so the resulting estimated marginal variances satisfy (\ref{eq:sigmaFOCs}) and are equal to $\hat \sigma^2(\Delta)$. 

Penalising the off-diagonal of $\Theta$ therefore results in the global MLE of $\sigma^2$, whereas penalising $\Delta$ results in the MLE conditional on the estimated $\Delta$. Since the likelihood for $\sigma^2$ can vary greatly for different fixed $\Delta$, the second option of penalising $\Delta$ seems preferable.

\section{Singular $S$}\label{sec:Singular}

The previous two sections show the advantages of using penalised likelihoods on $\Delta$
\begin{equation}\label{eq:DeltaPL}
    \argmax_{\theta \succ 0,\Delta \succ 0} \; l(\theta,\Delta \mid S) - \sum_{i<j} \pen(\Delta_{ij})
\end{equation}
On top of being scale invariant, they also better estimate graphs with high order nodes and estimate $\sigma^2$ by the MLE given the estimated $\Delta$. 

However, without further penalisation of other parameters, the solution to (\ref{eq:DeltaPL}) does not exist when $S$ is not positive definite, because the likelihood diverges along a path with $\lVert \theta \rVert \to \infty$ and $\Delta$ converging to a singular matrix. The penalty on $\Delta$ cannot counteract this, even with $\pen(\Delta_{ij}) \to \infty$ as $\abs{\Delta_{ij}} \to 1$, because there are singular $\Delta$ with all $\Delta_{ij} \in (-1,1)$. See \cite{carter2025existence2} for a thorough analysis of the existence of penalised likelihood estimates for GGMs. Instead, one must either use a non-separable penalty on $\Delta$ which actively penalises matrices with small eigenvalues, or the direction $\lVert\theta\rVert \to \infty$ must be penalised.



It was shown in \citet{carter2024partial} that a logarithmic penalty on $\theta$ maintains scale invariance of the selected GGM and in \citet{carter2025existence} that a sufficiently strong logarithmic penalty is enough to ensure existence of the penalised likelihood estimate for singular $S$. However, as with diagonal penalisation for the precision matrix, penalisation of $\theta$ results in inflated estimates of the marginal variances.

Instead existence of the penalised likelihood estimate may be achieved through penalisation of $\psi$. The resulting penalised likelihood estimate is
$$\argmax_{\sigma^2 \succ 0,\Delta \succ 0} \; l(\sigma^2,\Delta \mid S) - \sum_{i=1}^p \pen_D(\psi_i) - \sum_{i<j} \pen(\Delta_{ij}).$$
Because there is no penalisation of $\sigma^2$, it will be estimated via the conditional MLE $\hat \sigma^2(\Delta)$.

Since $\psi$ is the diagonal of $\Delta^{-1}$ and $\Delta$ has minimum eigenvalue tending to 0 if and only if a diagonal entry of $\Delta^{-1}$ goes to $\infty$, the direction $\psi_i \to \infty$ exactly characterises how $l(\sigma^2,\Delta \mid S) \to \infty$. Hence, sufficient penalisation of $\psi$ is enough to ensure existence of the penalised likelihood estimate.  Letting $m_0$ be the number of eigenvalues equal to 0 in $S$, as a corollary to \citet{carter2025existence} Theorem 1.1, I conjecture that the solution exists as long as 
$$ \liminf_{\psi_i \to \infty} \frac{\pen_{D}(\psi_i)}{\log \psi_i} > m_0/p.$$
Although penalisation of $\psi$ results in penalisation of the total dependence of $X_i$ on $X_{-i}$, and therefore larger penalisation of high degree nodes, this is exactly what is required when $S$ is singular. This is because in the small sample size setting $n<p$, within the observed data $X_i$ is a linear function of $X_{-i}$, which is characterised by $\psi_i \to \infty$. If one wishes to keep the possibility of high degree nodes in the estimated graph, it therefore seems prudent to set the penalty on $\psi_i$ to be as small as possible while ensuring existence, i.e. a logarithmic penalty. However, penalising $\psi$ instead of $\theta$ does give the possibility of moving away from logarithmic penalties while maintaining scale invariance of the estimator. This might be for computational reasons or because an estimated graph with low node degrees is desired, in which case a stronger penalty on $\psi$ can be used.

\section{Discussion}

The decomposition of $\Theta$ into partial correlations, VIFs and marginal variances, demonstrates some fundamental issues with penalised likelihood methods on $\Theta$. I believe this is enough evidence to redirect attention towards directly penalising partial correlations.

The main challenge for penalising $\Delta$ is in computation. This is because this parameterisation introduces non-convexity in the log-likelihood through the trace term. Progress has already been made on this front with two algorithms proposed for optimisation of PCGLASSO \citep{carter2025existence,bogdan2025identifying}. Further research is required for implementation of non-$L_1$ penalties on $\Delta$ and for applying a penalty directly to $\psi$ as recommended in Section \ref{sec:Singular}. The observation in Section \ref{subsec:Joint} that the trace term of the log-likelihood is constant when the marginal variances are estimated via $\hat \sigma^2(\Delta)$ is also an interesting avenue that could lead to computational advances.

Although this paper has focused on penalised likelihood estimates, the decomposition of $\Theta$ can be used to gain insight into any shrinkage method. For example, a number of papers reframe estimation of $\Theta$ through linear regression $X_i = \sum_{j \neq i} \beta_j^{(i)}X_j + \epsilon_i$. The regression coefficients and error variance can be written as \citep{meinshausen2006high}
\begin{align*}
    \beta_j^{(i)} = -\frac{\Theta_{ij}}{\Theta_{ii}} = -\Delta_{ij} \sqrt{\frac{\sigma_i^2 \psi_j}{\sigma_j^2 \psi_i}}, && \mathrm{Var}(\epsilon_i) = \frac{1}{\Theta_{ii}} = \frac{\sigma_i^2}{\psi_i}.
\end{align*}
Data can been standardised to remove the effect of $\sigma^2_i,\sigma^2_j$ and $\psi_i$ is common to all coefficients in this regression. So the important aspect is that $\Delta_{ij}$ is multiplied by $\sqrt{\psi_j}$. Penalised likelihood methods will therefore place higher penalty on $\beta_j^{(i)}$ associated to large $\psi_j$, i.e. when $X_j$ has high dependence on the other predictors. An interesting recent method of this style that aims to mitigate the effects of scaling is \cite{nguyen2026large}.

Another important application of this decomposition of $\Theta$ is in the Bayesian setting. The current standard is to use a marginal exponential prior on the diagonals $\theta$, but an exponential has a heavy lower tail and so will tend to encourage smaller values of $\theta$. As was shown, this can lead to smaller node degrees and poor estimation of $\sigma^2$. Additionally, an important aspect of the Bayesian framework is the ability to incorporate prior information. By separating the concepts of scale and dependence by $(\sigma^2,\Delta)$, this task is made easier.

\section*{Acknowledgments}

This work was supported by the EUTOPIA Science and Innovation Fellowship Programme and funded by the European Union Horizon 2020 programme under the Marie Skłodowska-Curie grant agreement No 945380.


\bibliography{references}

\appendix

\end{document}